\documentclass[twocolumn,showpacs,preprintnumbers,amsmath,amssymb]{revtex4-1}

\usepackage{graphicx} 
\usepackage{dcolumn} 
\usepackage{bm} 

\begin{document}

\title{\large Tunable Non-local Coupling between Kondo Impurities}

\author{D. Tutuc $^{1}$$^{\ast}$}
\author{B. Popescu $^{1}$$^{\dag}$}
\author{D. Schuh $^{2}$}
\author{W. Wegscheider $^{2}$$^\S$}
\author{R.~J. Haug $^{1}$}

\affiliation{$^{1}$ Institut f\"ur Festk\"orperphysik, Leibniz
Universit\"at Hannover, Appelstr. 2, 30167 Hannover, Germany}
\affiliation{$^{2}$ Institut f\"ur Experimentelle und Angewandte
Physik, Universit\"at Regensburg, Universit\"atsstrasse 31, 93053
Regensburg, Germany}
\date{\today}

\begin{abstract}
We study the tuning mechanisms of Ruderman-Kittel-Kasuya-Yosida (RKKY) exchange interaction between two
lateral quantum dots in the Kondo regime. At zero magnetic field
we observe the expected splitting of the Kondo resonance and estimate the
non-local coupling strength as a function of the asymmetry between
the two Kondo temperatures. At finite magnetic fields a chiral
coupling between the quantum dots is observed in the Kondo
chessboard and we probe the presence of the exchange interaction
by analyzing the Kondo temperature with magnetic field.
\end{abstract}

 \maketitle


The high versatility and tunability of semiconductor quantum dots (QDs) have attracted increasing interest in the past years and their investigation as magnetic impurities has allowed even the realisation of the artificial Kondo system~\cite{Goldhaber-Gordon-98,Goldhaber-PRL-98,Cronenwett-98}. Kondo effect is a signature of spin entanglement in a
many-electron system, where delocalized electrons screen a
localized spin, leading to the formation of a new singlet ground
state. The main signature of Kondo effect is the formation of a
peak in the density of states at the impurity site due to
successive spin-flips at temperatures below $T_{K}$, the so-called \emph{Kondo temperature}, essentially the energy scale
describing the binding energy of the spin singlet formed between
the localized, unpaired electron and delocalized electrons in the
leads.


Kondo impurities also interact with one another via carrier
mediated spin-spin interactions that compete with the local interactions that give rise to Kondo effect. The competition between the two effects is a form of quantum entanglement between two or more spins, usually discussed in the framework of the \emph{two-impurity Kondo problem}~\cite{Hewson-93}. The Ruderman-Kittel-Kasuya-Yosida (RKKY) is an example of such a carrier mediated interaction~\cite{Hewson-93}. In quantum dot
systems, even if the observation of the RKKY-Kondo competition has
been previously claimed~\cite{Craig-04,Sasaki-06} a conclusive
understanding is missing~\cite{Simon-05,Vavilov-05}, and the observed results have also been
explained in terms of a Fano antiresonance~\cite{Martins-06}.
While the description in terms of Fano antiresonance is
suitable for high interdot hopping, the explanation in terms
of RKKY is more appropriate in the limit of strongly localized electrons. In between, the two effects can coexist~\cite{Martinek-10,Zitko-10}. Due to the fact that the RKKY interaction provides a way for spin entanglement control beyond the nearest-neighbor restraint, and its strong dependence on the Fermi
wavevector, the implications to quantum
information processing cannot be underestimated~\cite{Loss-98}.

 \begin{figure}
  \includegraphics[scale=0.8]{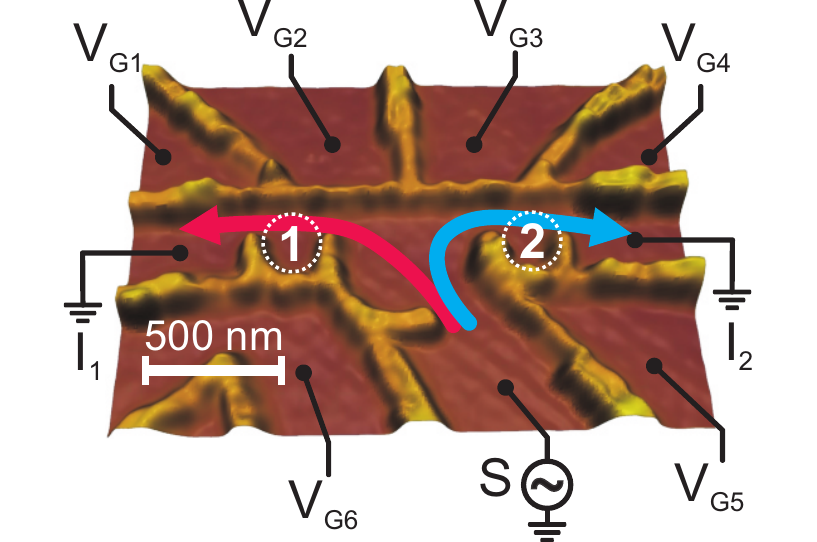}
  \caption{AFM image of our device defined by oxide lines
  (bright/yellow). Quantum dots, 1 and 2, are connected to
  a common source S, and each to individual
  drains. Six in-plane gates, G1 to G6, control the potentials
  of the dots and coupling to the leads. The arrows mark
  the measured transport paths.}
  \label{fig1}
 \end{figure}

In this letter we study the tuning mechanisms of the RKKY exchange
interaction strength in a system with two lateral QDs
coupled to a central open conducting region~\cite{Craig-04}. As
mentioned before, $T_{K}$ describes the binding
energy of the spin singlet formed between the confined electron
spin and the leads, with an analytical form given by
$T_{K1(2)}\sim~$exp$(-1/(2D_{C}K_{1(2)}))$~\cite{Hewson-93}, where
$T_{K1(2)}$ is the Kondo temperature of quantum dot 1(2), $D_{C}$
the density of states (DOS) in the leads, and $K_{1(2)}$ the Kondo
coupling to the leads. Assuming a continuous DOS in the central
region, the RKKY exchange interaction strength is given by $J\sim
K_{1}K_{2}$~\cite{Simon-05}. Therefore by tuning the tunnel coupling between the quantum dot and the
leads, the Kondo coupling $K_{i}$ is also tuned, and with it the
strength of the exchange interaction. Using this method we study
how the exchange interaction strength $J$ changes by tuning the
system into an asymmetric state, more explicitly by tuning the
coupling of QD1 to the central region with respect to
QD2. Furthermore, a perpendicular magnetic field induces
Landau levels and edge states. At high fields the
edge states in the leads remain partially spin unpolarized, so Kondo effect can still be observed in
transport~\cite{Keller-01, Fuehner-02, Stopa-03, Kupidura-06}.



Our device was produced by local anodic oxidation with an
AFM~\cite{Ishii-95,Held-98,Keyser-00}, on a
GaAs/Al$_{x}$Ga$_{1-x}$As heterostructure containing a
two-dimensional electron gas (2DEG) 37~nm below the surface
(electron density $n_{e}=3.95\times10^{15}~m^{-2}$, and mobility
$\mu=65~m^{2}/Vs$ measured at 4.2~K). The structure consists of
two quantum dots~(Fig.~\ref{fig1}), QD1 and QD2, connected
each to an individual drain, and to a central common reservoir,
further coupled to the Source reservoir $S$ via an 1D
constriction. The distance between the QDs is about $600$~nm.
Using standard lock-in technique we measure the differential
conductances $g_{1}=dI_{1}/dV_{S}$ and $g_{2}=dI_{2}/dV_{S}$
across each dot in parallel by applying an ac excitation of
$10~\mu$V with a frequency of $83.3$~Hz at the source S. The measurements were
performed in a $^{3}$He/$^{4}$He dilution refrigerator with an
electron temperature of $\sim80$mK. Source-drain bias dependencies
revealed a charging energy of \hbox{$0.25$}~meV for QD1,
\hbox{$0.3$}~meV for QD2 and single energy level spacings of
$\Delta\approx 50$~$\mu$eV. Gates G2 and G3 are used
to control the coupling of the two dots to the central region,
while G5 and G6 are used as plunger gates.

 \begin{figure}
  \includegraphics{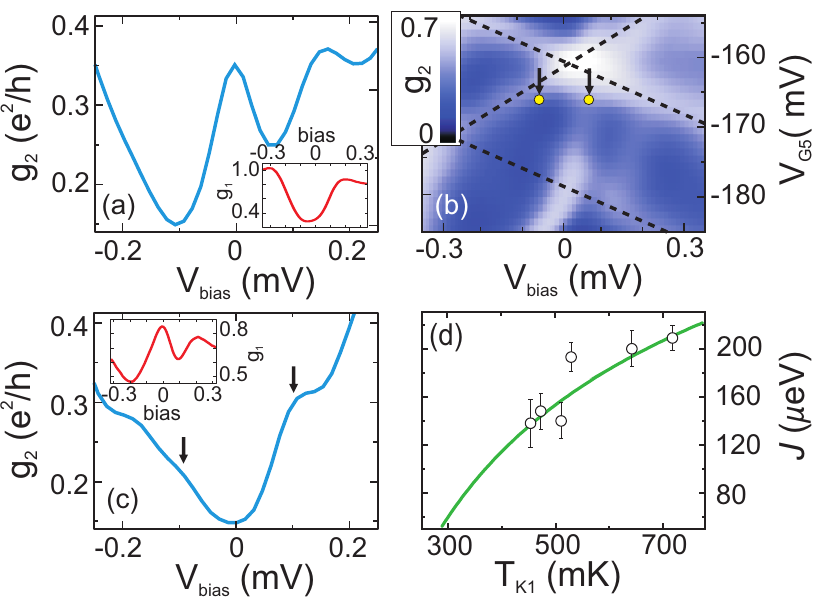}
  \caption{(color)(a) Conductance $g_{2}$ through QD2 showing a zero-bias anomaly, when QD1 is tuned in a non-Kondo valley (inset).
  (b) Color scale plot of $g_{2}$ as a function of gate G5 and bias voltage, showing the splitting of the QD2 ZBA. Dot+arrow mark the splitting.
  (c) Conductance $g_{2}$ showing a split zero-bias anomaly when QD1 is tuned in a Kondo
  state (inset) taken along the dashed line in (b). The black
  arrows mark the position of the split ZBA.
  (d) Exchange interaction strength $J$ as a function of $T_{K1}$, fit to $1/ln(x)$ (line).}
  \label{fig2}
 \end{figure}


Both quantum dots can be individually tuned in the Kondo regime
and, by changing the electron number one by one, we can switch
from a Kondo to a non-Kondo valley. Figure~\ref{fig2}(a) shows a
Kondo resonance of QD2, or the so-called zero-bias anomaly (ZBA),
as a function of dc bias, while the inset shows the corresponding
situation in QD1, i.e. tuned to the middle of a Coulomb valley
that does not exhibit a Kondo resonance. To estimate the Kondo temperature we use $T_{K}=(w\pi\delta)/(4k_{B})$~\cite{Hewson-93}, where $w$ is the Wilson number, $k_{B}$ the Bolzman contant and $\delta$ the half width at half maximum of the Kondo resonance~\cite{Comment-02}. A value of 280~mK is obtained for $T_{K2}$. Changing the electron number in
QD1 by one ($V_{G6}$ from -37 to -20~mV) brings it into a Kondo
state (inset of Fig.~\ref{fig2}(c)), characterized by a Kondo
temperature $T_{K1}\simeq720$~mK \cite{Comment-01} for
$V_{G2}=-48$~mV. The color plot of the differential conductance
$g_{2}$ through QD2 as a function of bias voltage and gate $G5$
(Fig.~\ref{fig2}(b)) shows the behavior of the QD2 ZBA when QD1 is
tuned in the middle of the Kondo valley for $V_{G2}=-54$~mV. From $V_{G5}\cong-175$~mV the ZBA starts to deviate to the right along with a decrease in conductance, that suggests the beginning of a splitting. At $V_{G5}\simeq-167$mV beside the Kondo peak at $\sim75$$\mu$V bias, a shoulder emerges at $\sim-70$$\mu$V (marked in
Fig.~\ref{fig2}(b) by two dots and arrows). The asymmetric peak splitting is interpreted in terms of coupling asymmetry, QD2 being stronger coupled to the central region. The fact that we see a dependence on the Kondo energy scale indicates that the splitting is caused by the RKKY exchange interaction. The splitting of
the Kondo resonance gives direct access to $J$ using the relation
$eV_{bias}=\pm J/2$ (black arrows in Fig.~\ref{fig2}(c))~\cite{Aguado-00,Lopez-02,Simon-05} and the
obtained value is about $200$~$\mu$eV. However the ZBA splitting
is observed only in QD2, while in QD1 we see just a suppression of
the Kondo resonance, probably due to the strong asymmetry between the two Kondo scales. By adjusting $V_{G2}$ between -42 mV to -56 mV
the coupling of QD1 to the central region is tuned, and the Kondo
temperature of QD1 at the same time. Therefore we can measure
the magnitude of the splitting as a function of $T_{K1}$
(Fig.~\ref{fig2}(c)). The analytical relation for the Kondo
coupling $K_{i}\sim-1/(2D_{C}ln(T_{Ki}))$ gives an asymptotic
behavior of the RKKY exchange interaction strength $J$ as a
function of the Kondo temperature $T_{K}$, and from a fit to $J=a+b/ln(T_{K1})$ (line in Fig.~\ref{fig2}(d))\cite{Comment-03}, we obtain $a\sim1.2$~meV and $b\sim-6.5$~meV,  where $a$ is extracted as the maximum bandwidth of the central region~\cite{Simon-05}, while $b$ contains information about the density of states in the leads.


Kondo effect can be observed not only at zero magnetic field, but
also at finite perpendicular magnetic field. Since one of its main
conditions is the presence of both spin up and spin down in the
leads, the observation of Kondo effect in magnetic field requires
largely unpolarized leads. In our sample we can observe Kondo
effect up to 5.5~T in both QDs. This corresponds to a filling
factor of the leads $\nu_{leads}\simeq3$ (with $\nu=n_{e}h/eB$,
where $n_{e}$ is the electron density, $h$ is the Planck constant,
$e$ the elementary charge and $B$ the magnetic field), showing that the leads are not fully polarized. In the same
magnetic field range, the quantum dot filling factors are lower,
i.e. $\nu_{QD}<\nu_{leads}$, and basically their properties are
governed by the lowest Landau level (LL0) formed at the edge and
Landau level 1 (LL1) formed at the core of the dots. Because the
tunnel coupling of the core to the leads is not high enough, Kondo
effect  in magnetic field involves only transport through the edge
of the dot, which is closer to the leads. By changing the total
spin of the edge by $\frac{1}{2}$ one can change Kondo transport
through the dot. The total spin number in the edge is given by the
number of electrons and it can be changed in two ways: either we
add an electron from the leads or from the core of the dot. The
first mechanism is the result of increased voltage on a nearby
plunger gate, increasing also the total electron number by one,
while the second mechanism is the result of adding one flux
quantum to the dot by increasing the magnetic field, and
redistributing an electron from the core to the
edge~\cite{Fuehner-02}.

 \begin{figure}[t!]
  \includegraphics{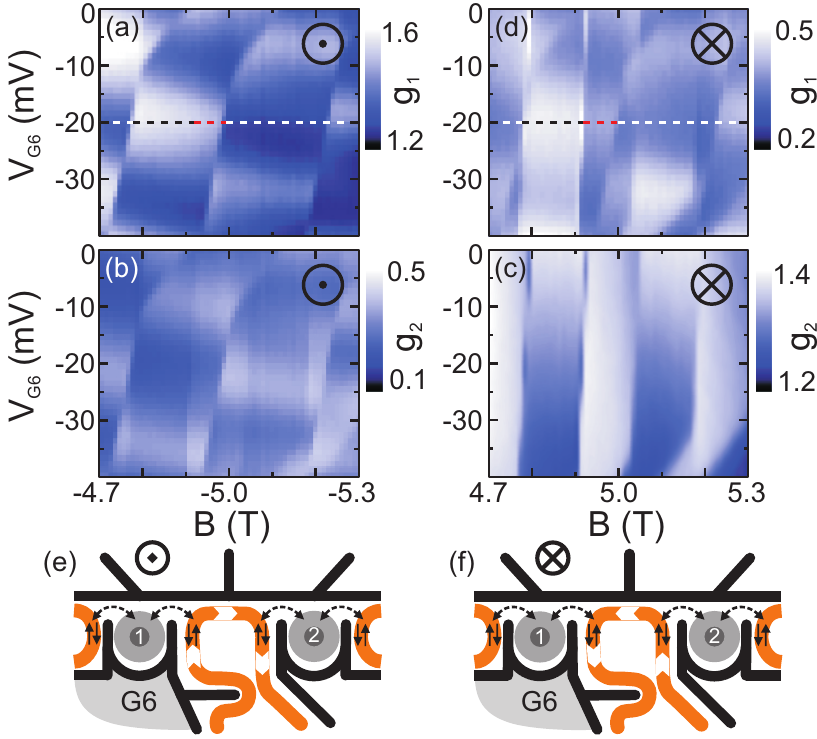}
 \caption{Differential conductance $g_{1}$ through QD1 as a function
 of $V_{G6}$ and magnetic field, for negative (a) and positive (d) magnetic field polarity.
 Differential conductance $g_{2}$ through QD2 as a function of $V_{G6}$ and
 B measured for negative (b), and positive (c) magnetic field polarity, measured simultaneously with (a), and (d) respectively. (e)-(f) Schematic diagram
 of the sample in magnetic field: with black - AFM oxide lines, orange - edge states. Landau Level 1 (dark grey) and Landau level 0 (light grey) are marked in the QDs.}
  \label{fig3}
 \end{figure}


The result of these two mechanisms is a Kondo effect modulation as
a function of gate voltage and magnetic field in a regular pattern
of high-low differential conductance, generally referred to as the
Kondo chessboard~\cite{Keller-01, Fuehner-02, Stopa-03,
Kupidura-06, Rogge-06}. In Fig.~\ref{fig3}(a) there is shown the expected
Kondo chessboard pattern exhibited by QD1, with alternating
regions of high (bright) and low (dark) differential conductance
as a function of the voltage applied on G6 and magnetic field. In
contrast, QD2 exhibits a more complicated pattern
(Fig.~\ref{fig3}(b)). By changing the direction of the magnetic
field the situation is reversed. Now we observe a regular
chessboard pattern in the transport through QD2
(Fig.~\ref{fig3}(c)) - here the total electron number remains
unchanged, because gate G6 has a very small lever arm on QD2, and
one sees only a modulation of the conductance by the magnetic
field. At the same time QD1 (Fig.~\ref{fig3}(d)) exhibits a
pattern similar to the one in Fig.~\ref{fig3}(b). The data
presented in Fig.~\ref{fig3} (a) and (b), respectively (c) and
(d), are acquired in parallel, in the same range of $V_{G6}$ and
B-field, so the differences arise only from changing the polarity
of the perpendicular magnetic field, which in fact changes the
direction of electron transport in the edge states formed in the
central region between the dots. In the first case
(Fig.~\ref{fig3}(a) and (b)), the edge state picture corresponds
to the one depicted in Fig.~\ref{fig3}(e), that is the edge state
transport direction is clockwise in the central region, therefore
transport through QD2 will be sensitive to potential changes in
the edge states generated by electron transport through QD1, i.e.
the sample will behave as a current divider. As a consequence QD2
will exhibit a combination of its own chessboard pattern and a
negative of the QD1 pattern (Fig.~\ref{fig3}(b)). In the second
situation (Fig.~\ref{fig3}(c) and (d)) the edge state transport
direction is reversed - as depicted in Fig.~\ref{fig3}(f), i.e.
the electrons move counter-clockwise in the edge states in the
central region. Hence, now QD1 will exhibit a combination of both
chessboard patterns (Fig.~\ref{fig3}(d)). The measurements in
Fig.~\ref{fig3}(a)-(d) demonstrate a chiral coupling between the
quantum dots via the edge states formed in the central region.
Using these conductance plots one can identify the regions were
only one or both quantum dots exhibit transport through a Kondo
state. Along the dashed lines in Fig.~\ref{fig3}(a) and (d) two
such situations are marked, that is between $\pm4.77$~T and
$\pm4.97$~T (black+red part on the dashed line) QD1 exhibits
transport through a Kondo state. In the $\pm4.9$~T to $\pm5.04$~T
interval QD2 also exhibits Kondo
transport (Fig.~\ref{fig3}(c)), therefore from $\pm4.9$~T to
$\pm4.97$~T (red part on the dashed lines) both quantum dots are in the Kondo regime. It is worth
mentioning that the sum of the differential conductances through
the two dots does not change with the magnetic field direction,
hence a departure from the Onsager symmetry
relations~\cite{Onsager-31} is unlikely.

 \begin{figure}[b!]
  \includegraphics{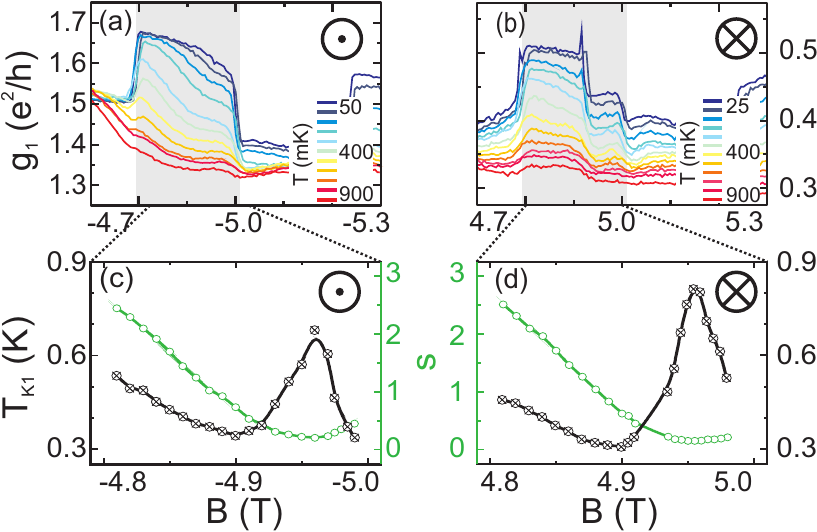}
 \caption{Temperature dependence of QD1 for negative (a) and positive (b) magnetic field polarity,
 taken along the white dashed lines in Fig.~\ref{fig3}(a), and \ref{fig3}(d) respectively.
 $T_{K1}$ ($\otimes$) versus magnetic field for negative (c), and positive (d) polarity; with green($\circ$) the $s$ fitting parameter.}
  \label{fig4}
 \end{figure}


In order to probe the exchange interaction between the two quantum
dots in perpendicular magnetic field in detail we investigate the
temperature dependence of QD1 taken along the dashed lines
in Fig.~\ref{fig3}(a) and \ref{fig3}(d). In Fig.~\ref{fig4}(a) the
sharp steps in the differential conductance at $-4.77$~T and
$-4.97$~T mark the onset of Kondo effect in QD1, and it can be
observed that around $-4.9$~T there is a change in the temperature
dependence behavior, suggesting the presence of a second energy
scale. For the opposite polarity (Fig.~\ref{fig4}(b)), due to the
change in the edge state chirality, an extra step in the
conductance of QD1 marks the onset of Kondo effect in QD2 at
$\sim4.9$~T. Using the relation $ g(T)=g_{0}
(T_{K}^{'}/(T^{2}+T_{K}^{'2}))^{s}$ with
$T_{K}^{'}=T_{K}/(\sqrt{2^{1/s}-1})$~\cite{Goldhaber-PRL-98} to
fit the measured temperature dependence within the $\pm4.77$~T to
$\pm4.97$~T interval, we obtain a change in $T_{K1}$ with
magnetic field (Fig.~\ref{fig4}(c) and \ref{fig4}(d)). From
$\pm4.7$~T up to approximatively $\pm4.9$~T, $T_{K1}$ shows a
monotonic decrease, which is followed by an increase up to
$\sim\pm4.95$~T, then it decreases rapidly. The position where
$T_{K1}$ starts to increase coincides with the onset of Kondo
effect in QD2. The fitting procedure is done with the $s$
parameter left free and the obtained values are shown along with $T_{K1}$ in
Fig.~\ref{fig4}(c) and (d).


As mentioned
before, at high perpendicular magnetic fields Kondo effect
involves transport only through Landau Level 0 (LL0) energy
states. Since the measurement is done at fixed gate voltage, i.e.
at fixed energy, as the magnetic field is increased, the ground
state moves lower in energy and the electron wavefunction is compressed, thus the Kondo coupling
is reduced, which explains the initial decrease of $T_{K1}$. In
this region $s$ shows a similar decrease with magnetic field, with
values between 2.5 and 0.5 which indicate a clear departure from
the $0.22$ value for spin $\frac{1}{2}$~\cite{Goldhaber-PRL-98}.
However similar values have been previously reported in the Kondo
chessboard~\cite{Fuehner-02}. On the other hand, the increase and
decrease of $T_{K1}$ around $\sim\pm4.95$~T, which is correlated
with the onset of Kondo effect in the other quantum dot, cannot be
explained by single dot physics, although in this region $s$ is
remarkably close to the $0.22$ value. Therefore non-local exchange
interaction has to be considered. The zero-bias temperature
dependence of a split ZBA (either by magnetic field or by RKKY)
follows that of the zero magnetic field Kondo resonance at high
temperatures \cite{Costi-00, Martinek-10} and a change in the ZBA
splitting will appear as a change in the Kondo temperature.
Therefore the $T_{K1}$ peaks seen in Fig.~\ref{fig4}(c) and
\ref{fig4}(d) are probably not due to an actual increase in Kondo coupling,
but are attributed to a change in ZBA splitting in the presence of
the exchange interaction between the spins of the dots. Even
though by reversing the magnetic field direction the electron
transport direction is changed, the general behavior of $T_{K1}$
is very similar for both magnetic field directions
(Fig.~\ref{fig4}(c)-(d)), questioning whether the chirality of the
edge states influences also the exchange interaction between the
spins of the dots.

In conclusion, we have investigated the exchange interaction
between the two quantum dots as a function of Kondo temperature
asymmetry and perpendicular magnetic field. At $B=0$ we find the
expected $\sim1/ln(T_{K})$ dependence of the RKKY interaction
strength $J$ on the Kondo energy scale. At finite magnetic fields
we observe a chiral coupling between the quantum dots in the Kondo
regime and we probe the presence of the RKKY exchange interaction
by a Kondo temperature analysis.

The authors would like to thank N. Ubbelohde, C. Fricke and F.
Hohls for their help with the measurement setup and data analysis,
as well as E. R\"{a}s\"{a}nen, J. Martinek, R. Zitko and R. Lopez
for many valuable discussions. We acknowledge financial support
from the German Excellence Initiative via QUEST and the NTH School
for Contacts in Nanosystems.


$^{\ast}$e-mail: tutuc@nano.uni-hannover.de; $^\dag$present address:  School of Engineering and Science, Jacobs
University Bremen, Campus Ring 1, 28759 Bremen,
 Germany;
 $^\S$present address: Laboratorium f\"ur Festk\"orperphysik, ETH Z\"urich, Schafmattstr. 16, 8093 Z\"urich




\begin{thebibliography}{20}
\bibitem[{\citenamefont{}(2002)}]{Comment-02} The HWHM of the ZBA is obtained from a Lorenz fit with 0 background.
\bibitem[{\citenamefont{}(2002)}]{Comment-01} The Kondo temperature of each quantum dot was measured while the other dot was not in a Kondo state.
\bibitem[{\citenamefont{}(2002)}]{Comment-03} Due to limited data range, a linear fit would work as well.
\end{thebibliography}



\end{document}